\documentstyle[sprocl]{article}

\input{psfig}

\def\Journal#1#2#3#4{{#1} {\bf #2}, #3 (#4)}

\def\NPB{{\em Nucl. Phys.} B}
\def\PLB{{\em Phys. Lett.}  B}
\def\PRL{\em Phys. Rev. Lett.}
\def\PRD{{\em Phys. Rev.} D}

\def\be{\begin{equation}}
\def\en{\end{equation}}
\def\bea{\begin{eqnarray}}
\def\ena{\end{eqnarray}}
\def\hbo{\hbox to 1cm {\hfill }} 
\def\tr{\hbox {tr}} 
\def\fm{\hbox {fm}} 

\begin{document}

\title{ CONFINEMENT AND THE VORTEX VACUUM OF SU(2) LATTICE GAUGE THEORY }

\author{ Kurt Langfeld }

\address{ Institut f\"ur Theoretische Physik, Universit\"at T\"ubingen, \\ 
Auf der Morgenstelle 14, D-72076 T\"ubingen, Germany \\ 
E-mail: langfeld@alpha6.tphys.physik.uni-tuebingen.de } 


\maketitle\abstracts{ The vortex theory which emerges 
from SU(2) lattice gauge theory by center projection is briefly 
reviewed. In this vortex picture, quark confinement is due to 
percolating (closed) vortices which are randomly linked to the 
Wilson loop. The deconfinement phase transition appears as a 
de-percolation phase transition. 
}

\section{Introductory remarks}\label{sec:1}

Considering the nuclear force between hadrons as residuum 
of a confining force between the constituents of the hadrons explains 
why the understanding of quark confinement acquires highest priority 
in modern medium energy physics. In this context, 
lattice gauge theory~\cite{rothe} provides a convenient tool since it 
covers the non-perturbative aspects of quantum field theory which 
are important for the low energy regime. 
Numerical simulations of lattice QCD allow for a comparison of theoretical 
predictions with experimental data  and trigger the interpretation 
of experimental data. Numerical simulations of pure SU(2) Yang-Mills 
are also very helpful to 
get insights into the basic mechanisms of gluon dynamics, which presumably 
dictates the structure of hadrons. The finite amount of computational power 
is the only limiting factor of the accuracy of the theoretical predictions. 
Modern computers allow for a number lattice points which corresponds 
to an ultra-violet cutoff of several GeVs by a size of the lattice 
universe of several fms. This size of the universe is several times 
of the size of the generic coherence length of the gluon sector, and 
the UV-cutoff is large enough to cover a wide span of low lying 
excitations. 

\bigskip 
Since lattice gauge simulations employ a discretized version of 
(Euclidean) space-time (with lattice spacing $a$), the crucial task 
is to extrapolate the lattice results to the continuum limit 
$a \rightarrow 0$. From continuum SU(2) gauge theory, we learn that 
the bare coupling constant $g$ is a definite function of the UV regulator 
$\Lambda $. In fact, the self-consistent treatment of the gauge theory 
at high energies reveals~\cite{yn83} 
\be 
\frac{1}{g^2(\Lambda )} \; = \; \frac{1}{g^2(\mu )} 
\; + \; \frac{11}{24 \pi ^2 } \, \ln \left( \Lambda ^2 / \mu ^2 \right) 
\; , \hbox to 3cm {\hfill (pure SU(2))\hfill} 
\label{eq:1} 
\en 
where the prefactor of the logarithm depends on the gauge group and 
on the number of degrees of freedom. 
Keeping in mind that the UV cutoff in lattice simulations is provided by 
the lattice spacing, i.e. $\Lambda := \pi /a $, and defining 
$\beta := 4/g^2$, we conclude from (\ref{eq:1}) that 
\be 
a^2(\beta ) \; \propto \; \exp \left\{ - \frac{ 6 \pi ^2 }{11} \, 
\beta \right\} \; , \hbox to 2cm {\hfill for \hfill} 
\beta \gg 1 \; . 
\label{eq:2} 
\en  
Any {\it physical } quantity which is measured in units of the lattice 
spacing $a$, e.g. $\sigma a^2 $ with $\sigma $ the string tension, 
must show a unique dependence on $\beta $ and exponentially 
decreases for large values of $\beta $. In particular, the ratio 
of two {\it physical } quantities, e.g. $m^2 a^2 / \sigma a^2 $ with 
$m$ a glue-ball mass, becomes independent of $\beta $ for $\beta \gg 1$ 
(renormalization group invariance), and one safely extrapolates the 
mass squared in units of the string tension to the continuum limit 
$a \rightarrow 0$. Due to $\beta $--independence, the string tension 
becomes the only parameter of pure Yang-Mills theory (dimensional 
transmutation). 

\bigskip 
In the recent past, many observables have been addressed in computer 
simulations. For a review see~\cite{simdata}. 
Besides these ''lattice measurement'' of observables, many efforts 
were devoted to support or to invalidate scenarios of quark 
confinement. Among the promising ideas, I would like to mention the 
scenario of the dual super-conductor~\cite{tho76} which assumes 
a condensation of Abelian monopoles which then yields 
a dual Meissner effect. Over the last two decades, the evidence 
has increased that a vortex type structure of the vacuum is responsible 
for quark confinement~\cite{mack,cornwall,tom93,faber,error}. 
In this paper, we will further pursue the vortex picture and 
will study its properties at finite temperatures.

\section{Towards the roots of confinement} 

The fundamental degrees of freedom of pure SU(2) lattice Yang-Mills 
theory are SU(2) matrices $U(b)$ which are defined at the links $b$ 
of the lattice. The partition function is a functional integral over 
these matrices, i.e. 
\be 
Z \; = \; \int {\cal D} U \; \exp \{ - S _{W} \} \; , \hbo 
S_W \; = \; \beta \sum _{p} \left[1 - {\cal P}(p) \right]  \; , 
\label{eq:3} 
\en 
where ${\cal P } := \prod _{b\in p} U(b) $ is defined at the plaquette 
$p$ of the lattice and ${\cal D}U$ includes the Haar measure. 
Despite the discretization of space-time, action and partition function 
enjoy an exact gauge invariance $U(b) \rightarrow \Omega (x) U(b)$, 
$\Omega \in SU(2) $. 
The Wilson action $S_w$ reduces to the standard action of continuum 
Yang-Mills theory in the naive continuum limit $a \rightarrow 0$. 
Other choices are possible and differ from $S_w$ by {\it irrelevant 
terms }. These irrelevant terms can be either chosen to improve 
the convergence towards the continuum limit~\cite{perfectactions} or 
to suppress the statistical noise~\cite{noisesuppress}. 

\bigskip 
In this paper, we will focus on the mechanism of quark confinement. 
The potential $V(r)$ of two static quarks, located at distance $r$, 
can be obtained from the Wilson loop ${\cal W}$
\be 
{\cal W} := \prod _{b\in {\cal C}} U(b) \; , \hbox to 1cm {\hfill } 
\langle {\cal W} \rangle \; \propto \; \exp \left\{ 
- V(r) T \right\} \; \; \; \; \; (T \rightarrow \infty)\; , 
\label{eq:4} 
\en 
where ${\cal C}$ is a rectangular of spatial extension $r$ and 
extension $T$ in time direction. The so-called {\it area law }, i.e. 
$\langle {\cal W} \rangle \propto \exp \{ - \sigma {\cal A} \}$ with 
${\cal A}$ the minimal area enclosed by ${\cal C}$, is considered as 
confinement criterium, since one observes a linear rising confining 
potential $V(r) = \sigma \, r $ in this case.

\bigskip 
In a recent important work~\cite{faber}, the authors introduced the so-called 
{\it center projection}, which considerably reduces the number of degrees 
of freedom, while it preserves those relevant for confinement. 
Center projection firstly exploits the gauge degree of freedom 
for maximizing $\sum _b \{ \tr U(b) \}^2$. After this gauge fixing, 
the link variables $U(b)$ are as close to $\pm 1$ as possible. 
Secondly, one projects $U(b)$ onto the corresponding $Z_2$ variable 
${\cal U}(b)$. One easily checks that the particular gauge invariance 
provided by $\Omega (x) \in Z_2$ is not affected by the projection 
technique. This implies that this technique induces a $Z_2$ gauge theory. 

\bigskip 
It turns out that the off-diagonal elements of the link matrices of 
a particular lattice 
configuration are generically not small implying that the 
(relative) error induced into a generic observable by projection 
can reach 60\%~\cite{error}. The crucial observation is that, by 
contrast, the string tension is almost unchanged~\cite{faber,error} 
(center dominance). 
This bears the conjecture that the induced $Z_2$ gauge theory still 
contains the degrees of freedom relevant for confinement. In the 
very recent past, many results support this idea~\cite{review}. 

\bigskip 
The $Z_2$ gauge theory can be considered as a vortex theory. 
One says that a vortex pierces the plaquette $p$ if 
$v(p):=\prod _{b \in p} {\cal U}(b) = -1$. In order for revealing the 
string type nature of the vortices for a given time slice, we consider 
the plaquettes constituting a cube $c$, part of the spatial hypercube 
of space-time, and find $\prod _{p \in c} v(p) = 1$. This implies 
that the number of vortices piercing the plaquettes of the cube $c$ must be 
even. For this reason, the vortices necessarily form closed lines in space.  

\bigskip 
Are these vortices lattice artifacts or do they survive the 
continuum limit? In order for answering this question, we investigated 
the vortex (area) density $\rho $ which counts the average number of 
vortices piercing an area element~\cite{error}. The important 
observation is that $\rho a^2$ exhibits the characteristic dependence 
on $\beta $ (see (\ref{eq:2})) signaling that the vortices are physical 
objects rather than lattice artifatcs. We roughly find $\rho 
\approx 2 \, \fm ^{-2}$. We subsequently studied the vortex interactions 
by calculating the correlations between points where vortices intersect 
the plane~\cite{inter}.It turns out that this interaction also shows 
the desired renormalization group behavior attesting physical relevance. 
The vortex interaction is medium range attractive and possesses a range 
of $ \approx 0.4 \, $fm (see~\cite{inter}). 

\bigskip 
Let us neglect the inter vortex correlations for getting 
an idea of the confinement mechanism in the vortex picture. 
The expectation value of the Wilson loop is in this case~\cite{inter} 
\be 
\langle {\cal W} \rangle \; = \; \langle \prod _{b\in {\cal C}} 
{\cal U}(b) \rangle \; = \; \langle \prod _{p \in {\cal A} } 
v(p) \rangle \; = \; \sum _n (-1)^n \, P(n) \; , 
\label{eq:5} 
\en 
where ${\cal C} = \partial {\cal A}$ and $P(n)$ is the probability of 
finding $n$ vortices which are linked to ${\cal C}$. The sum in 
(\ref{eq:5}) can be easily evaluated for the case of the random vortex 
model~\cite{inter}. One recovers the desired area law $\langle {\cal W} 
\rangle = \exp \{ - 2 \rho {\cal A} \}$. The string tension is 
$\sigma _{rand} = 2 \rho \approx (400 \, \hbox{MeV})^2$, which is 
in good agreement with the exact value $\sigma = (440 \, \hbox{MeV})^2$.

\section{ Nature of the deconfinement phase transition } 

Once we have evolved a definite scenario of the confinement mechanism 
in the vortex picture (last section), the important question is 
whether the vortex vacuum correctly accounts for the deconfinement 
phase transition at finite temperature. It is known for a long time 
that SU(2) Yang-Mills theory undergoes a deconfinement 
phase transition at a critical temperature $T_c$. Choosing $\sigma 
= (440 \, \hbox{MeV})^2$ as reference scale, one finds 
$T_c \approx 210 \, $MeV for a pure SU(2) gauge theory, and $T_c \approx 
150 \, $MeV for the realistic case of lattice QCD~\cite{kanaya}. 

\bigskip 
Our numerical simulations 
show that, for a pure SU(2) gauge theory, the heavy quark potential 
is almost unchanged by center projection even at finite 
temperatures~\cite{temperature}. In particular, the correct transition 
temperature $T_c$ is found resorting to the induced $Z_2$ gauge theory. 

\begin{figure}[t]
\rule{5cm}{0.2mm}\hfill\rule{5cm}{0.2mm}
\psfig{figure=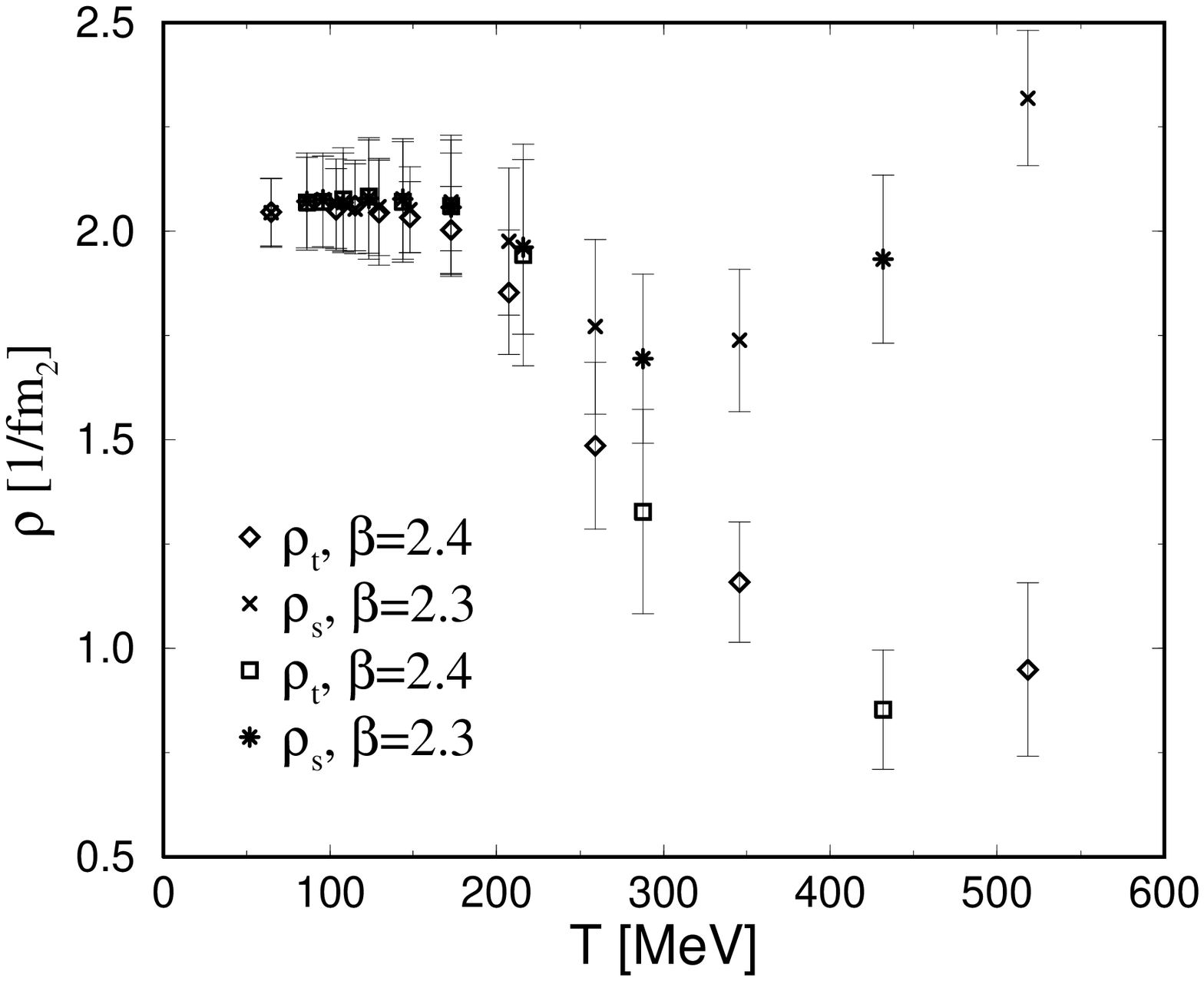,height=1.5in}
\hfill 
\psfig{figure=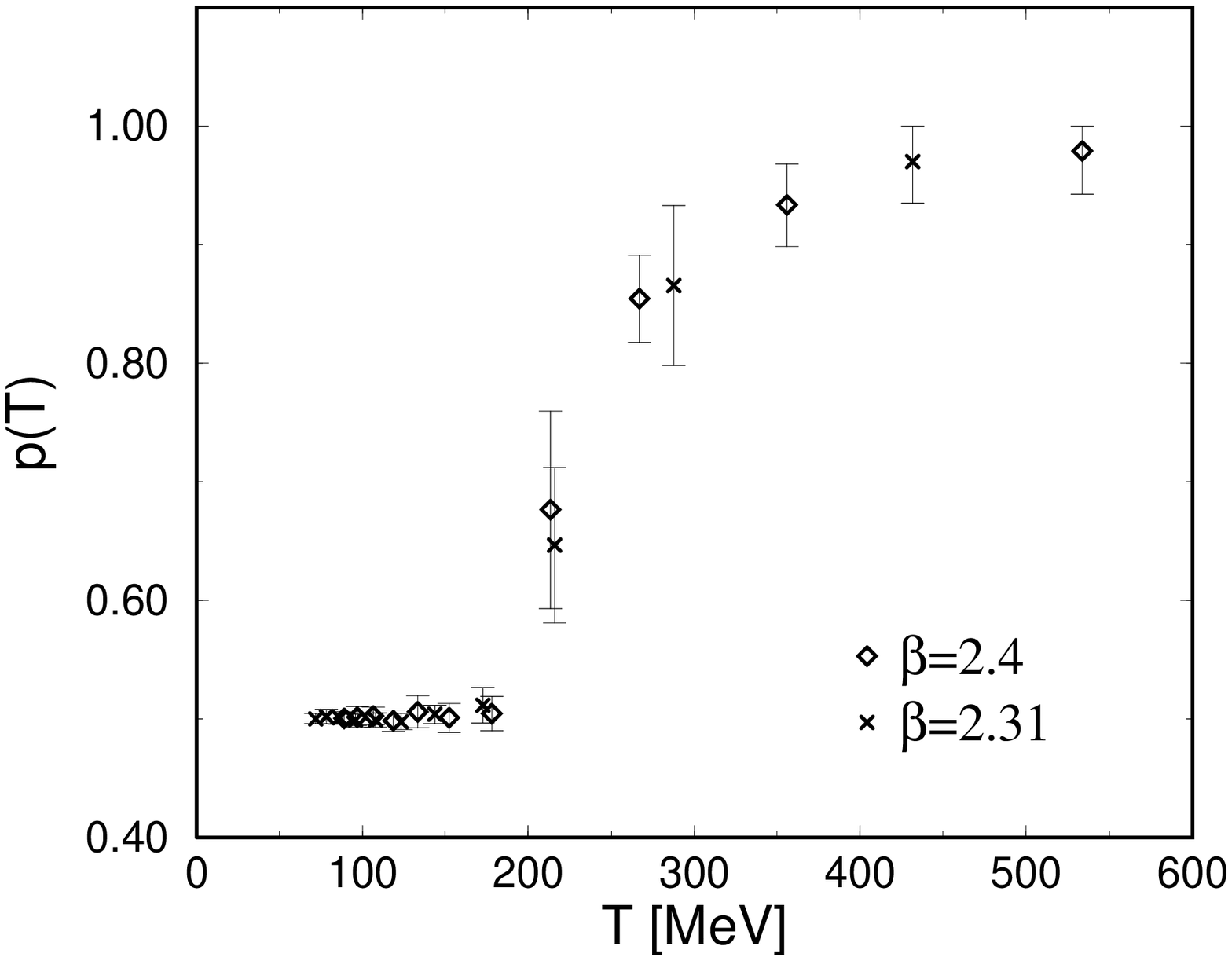,height=1.5in}
\rule{5cm}{0.2mm}\hfill\rule{5cm}{0.2mm}
\caption{ The timelike and spatial vortex area densities (left) and the 
vortex pairing (right) as function of temperature. 
\label{fig:1}}
\end{figure}

For revealing temperature effects, it is convenient to distinguish Wilson 
loops lying in a $tx$-plane from Wilson loops which are embedded in the 
spatial hypercube. in the following, we therefore contrast the time-like 
string tension to the spatial string tension. One should, however, 
keep in mind that the spatial string tension lacks a direct physical 
interpretation. 
In order for gaining definite insights into the nature of the 
deconfinement phase transition, we studied time-like and spatial 
vortex (area) densities as function of temperature, since these 
densities sets the scale of the corresponding string tensions. 
The result is shown in figure \ref{fig:1}. As expected, $\rho _s $ 
and $\rho _t$ coincide for $T< T_c$. For $T>T_c$, $\rho _t$ decreases, 
whereas $\rho _s$ is slightly suppressed at $T \approx T_c$ and 
strongly increases for $T \gg T_c$. The sum of $\rho _t $ and 
$\rho _s$ roughly stays constant. This indicates that the vortices 
get polarized along the time axis direction. However, these 
polarization cannot explain the sharp deconfinement phase transition, 
since $\rho _t$ has only dropped by a factor of $2$ at $T \approx 
2 T_c$. An inspection of (\ref{eq:5}) shows that the string tension 
would vanish if the number $n$ of vortices which pierce a given area 
is even. The quantity relevant for the phase transition is therefore 
the vortex pairing 
\be 
p (T) \; = \; \frac{ \langle  N _{even} \rangle }{ 
\langle  N _{even} +  N _{odd} \rangle } \; , 
\label{eq:6}  
\en 
where $N_{even(odd)}$ counts the events that an even (odd) number 
of vortices of a given MC configuration 
was linked to the Wilson loop. The numerical result 
for $p(T)$ is also presented in figure \ref{fig:1}. It shows a clear 
signal of the phase transition at $T=T_c$ (see~\cite{temperature}). 

\bigskip
The numerical results, presented above, are consistent with the 
following vortex picture: at low temperatures, i.e. $T< T_c$, 
the vortices {\it percolate}. Following the vortex network, one can 
''travel'' across the whole lattice universe. The infinite vortex cluster 
size provides the long range correlations which are imperative for 
observing the area law for large size Wilson loops. For $T >T_c$, the 
vortices stop percolating. In this case, the vortices form small size 
loops, which hardly intersect. Due to the small sizes of the vortex
clusters, only vortex clusters which are located close to the 
perimeter of the Wilson loop can contribute a non-trivial factor, i.e. 
$(-1)^n$, $n$ odd, and the de--percolation of the vortices at 
$T=T_c$ results in a perimeter law of the Wilson loop. In the 
vortex picture, the deconfinement phase transition appears as a 
de-percolation transition. Our most recent lattice calculations 
confirm this scenario~\cite{future}: the probability distribution 
of the maximum sizes of the vortex clusters reveals a clear 
signal of de-percolation at $T=T_c$.

\section{Conclusions} 

As already mentioned by 't~Hooft as long as twenty years ago, 
particular gauges might be more convenient than others for providing 
a definite picture of quark confinement. The so-called Abelian 
gauge support the intuitive picture of the dual Meissner 
effect~\cite{tho76}. Recent work~\cite{faber} proposes 
the so-called center gauge fixing and provides a definite prescription 
to project the SU(2) lattice Yang-Mills theory onto a $Z_2$ gauge theory. 
The latter theory offers a precise definition of a vortex theory, 
which is similar to those which were available in the literature for 
quite some time~\cite{mack,cornwall,tom93}. The vortex theory 
arising from center projection 
turned out to be very fruitful for understanding quark confinement. 
Let me summarize the most important results: the vortex theory 
reproduces the string tension within the statistical 
errors~\cite{faber,error} although the generic (relative) error produced 
by center projection is expected to be $\approx 60 \, $\%~\cite{error}. 
The vortex properties, such as the vortex area density or the inter vortex
correlation length, show the correct renormalization group 
behavior~\cite{error,inter}. The vortex picture extrapolates to the 
continuum limit. The vortices are physical objects rather than lattice 
artifacts. Zero temperature quark confinement is due to vortices 
which are randomly linked to the Wilson loop. The random distribution 
of the vortices is thereby due to vortex 
percolation~\cite{temperature,future}. 
The concept of vortex dominance of the string tension also extends 
to finite temperatures~\cite{temperature}. The deconfinement transition 
temperature is recovered in the effective 
$Z_2$ gauge theory to high accuracy. 
The deconfinement phase transition can be understood in the 
vortex theory as a de-percolation transition.

\section*{Acknowledgments}
It is a pleasure to thank my collaborators M.~Engelhardt, H.~Reinhardt 
and O.~Tennert.

\end{document}